# Experimental Realization of Two-Dimensional Buckled Lieb lattice

Haifeng Feng[1,2][†], Chen Liu[3][†], Si Zhou[1,4][*], Nan Gao[4], Qian Gao[5], Jincheng Zhuang[2], Xun Xu[1,2], Zhenpeng Hu[5], Jiaou Wang[3], Lan Chen[6,7][*], Jijun Zhao[4], Yi Du[1,2][*]

[1] Institute for Superconducting and Electronic Materials (ISEM), Australian Institute for Innovative Materials (AIIM), University of Wollongong, Wollongong, NSW 2500, Australia.
[2] BUAA-UOW Joint Research Centre and School of Physics, Beihang University, Beijing 100191, P. R. China.
[3] Beijing Synchrotron Radiation Facility, Institute of High Energy Physics, Chinese Academy of Sciences, Beijing 100049, P. R. China
[4] School of Physics, and Key Laboratory of Materials Modification by Laser, Ion and Electron Beams, Dalian University of Technology, Dalian 116024, China
[5] School of Physics, Nankai University, Tianjin 300071, P. R. China
[6] Institute of Physics, Chinese Academy of Science, Haidian District, Beijing 100080, China
[7] Songshan Lake Materials Laboratory, Dongguan, Guangdong 523808, China

* To whom correspondence should be addressed: yi_du@uow.edu.au (Y.D.), lchen@iphys.ac.cn (L.C.) and siz@uow.edu.au (S.Z.)
[†] These authors contributed equally to this work.

## Abstract

Two-dimensional (2D) materials with a Lieb lattice can host exotic electronic band structures. Such a system does not exist in nature, and it is also difficult to obtain in the laboratory due to its structural instability. Here, we experimentally realized a 2D system composed of a tin overlayer on an aluminum substrate by molecular beam epitaxy. The specific arrangement of Sn atoms on the Al(100) surface, which benefits from favorable interface interactions, forms a stabilized buckled Lieb lattice. Our theoretical calculations indicate a partially broken nodal line loop protected by its mirror reflection symmetry and a topologically nontrivial insulating state with a spin-orbital coupling (SOC) effect in the band structure of this Lieb lattice. The electronic structure of this system has also been experimentally characterized by scanning tunnelling spectroscopy and angle-resolved photoemmision spectroscopy. Our work provides an appealing method for constructing 2D quantum materials based on the Lieb lattice.

KEYWORDS: *Lieb lattice, geometry, spin-orbital coupling, 2D materials, nodal line, topplogical insultaing state*

The propagation of near-free electrons with unavoidable modulation by lattice geometry gives birth to distinct energy dispersions of electrons in solids. In particular, frustrated two-dimensional (2D) lattices often lead to interesting strongly correlated phenomena and the emergence of fascinating nontrivial structures[1-4], because of the reduced dielectric screening, pronounced quantum confinement effect, and promising quantum interference. Such a frustrated 2D lattice geometry is exemplified by the Lieb lattice, in which the atoms are arranged in an edge-centered square lattice with three atoms per unit cell (Fig. S1 in the Supporting Information (SI)). This geometry leads a unique electronic structure in the low energy regime, in which a flat band is sandwiched by two linear-dispersive bands (known as Dirac cones) at the M point of the Brillouin zone, if the electrons engage in nearest-neighbor and next-nearest-neighbor hopping. The Dirac cones give rise to massless carriers, while the flat band could enable realization of the fractional quantum Hall effect, superconductivity, and various topological nontrivial phases.[5-14] Its intriguing electronic structure associated with some exotic quantum states is of importance for both fundamental physics and practical applications, arousing intensive studies of the Lieb lattice in the past decade.

Due to its intrinsic structural instability, however, according to Landau-Peierls-Mermin-Wagner arguments, the Lieb lattice has only been previously achieved in some artificial systems such as molecular assemblies[15,16], photonic waveguide arrays[17-19], and cold atom systems[20]. The absence of any realization of stable Lieb lattice in inorganic materials has strongly hindered the development of the Lieb lattice in pursuing quantum effects. Very recently, the successful realization of certain monoelemental 2D materials (namely, Xenes) indicated that the specific epitaxial registry of foreign atoms on a suitable metal surface can stabilize the 2D system, which will result in a low-buckled 2D lattice.[21-25]

It is believed that a buckled Lieb lattice can be also formed epitaxially when certain criteria are met, as shown in Fig. 1a. There are two sublattices in this system, with the *A* atoms sitting at the four-fold hollow sites of the *B* sublattice. Intuitively, a feasible means to realize such a lattice is to adopt the (100) surface of a face-centered cubic (FCC) metal as *B* sublattice, while the *A* atoms are placed at the hollow sites of this metal substrate, forming a $\sqrt{2} \times \sqrt{2}$ structure with respect to the substrate. Nevertheless, the prerequisites should be a lower surface energy of the $\sqrt{2} \times \sqrt{2}$ structure compared with the bare substrate and sufficient strong interactions between the *A* atoms and nearest-neighbor *B* atoms, which prevent the formation of homogeneous clusters of *A* atoms on the surface.

Following the above criteria, we chose Sn as the *A* atoms to grow on an Al(100) surface,

because Sn and Al are immiscible with each other, which could prevent the formation of surface alloys. More importantly, with the large difference in electronegativity, Al (1.61 in Pauling scale) is expected to donate electrons to Sn (1.96),[26] suggesting sufficient strong interactions between them. After growth, a Sn overlayer on the Al(100) substrate with a $\sqrt{2} \times \sqrt{2}$ superstructure was successfully obtained. The scanning tunneling microscopy (STM) measurements confirmed the buckled Lieb lattice, and the angle-resolved photoemission spectroscopy (ARPES) measurements revealed the unique electronic structures. The theoretical calculations indicated that a partially broken nodal line (NL) loop protected by its mirror reflection symmetry exists in this peculiar lattice. Furthermore, an energy gap will be opened for the Dirac NL with inclusion of spin-orbital coupling (SOC), leading to a topologically nontrivial insulating state. This work provides an appealing way to search for novel 2D quantum materials and tune their electronic and topological properties by regulating the symmetry and the interaction between the crystal lattice and the substrate.

Figure 1a and 1b show an atomic model of the Sn adlayer on the Al(100) surface, in which the Sn atoms are located on the four-fold hollow sites of Al(100), forming a $\sqrt{2} \times \sqrt{2}$ superstructure. Our theoretical calculations reveal that the model is energetically favorable with negative binding energy ($E_b$) (as compared with bare Al(100) substrate) when the buckling height of Sn ($d$) ranges between 1.2 Å and 2.4 Å, as shown in Fig. 1c. On taking the optimal interlayer distance $d = 1.6$ Å, the Sn atoms are arranged in a square lattice with the nearest Sn−Sn distance of 4.0 Å. Obviously, the Sn adlayer and the topmost Al atoms form a buckled Lieb lattice. We also calculated the electronic band structure of this buckled Lieb lattice, as shown in Fig. 1d. An apparent nodal loop feature is visible at the energy of 1.9 eV below the Fermi level and centered at the $M$ point, resulting from the intersection of two bands from the Sn $p_z$ and Al $p_{xy}$ orbitals. One of the crossing points along $\Gamma$ to $M$ is opened by a small gap of about 0.1 eV. If we include the SOC in the calculations, additional 10 meV gaps are opened up at both nodes. Furthermore, we calculated the $Z_2$ topological invariant of the nodal line using the strategy proposed by Fu et al.,[27] and obtained a nonzero $Z_2 = 1$ (see SI for details), which implies that non-trivial topological fermions could exist in this buckled Lieb lattice. The NL was better revealed by the three-dimensional (3D) band structure in Fig. 1e and 1f, which takes the form of a rounded square, inheriting the $C_{4v}$ symmetry of this lattice system.

The electronic band structures of the buckled Lieb lattice with different buckling heights $d$ were also examined. For a planar Lieb lattice with $d = 0$, the complete NL is preserved, owing to the protection by mirror reflection symmetry (Fig. S3). When the Sn atoms are upper buckled, the Sn $p_z$ and Al $p_{xy}$ orbitals cross at the $M$ point, but the mirror reflection symmetry is lost with respect to the $M_{xy}$ plane. Consequently, an energy gap is opened at the crossing point along $\Gamma$ to $M$ by the hybridization of the Sn $p_z$ and Al $p_{xy}$ states, which varies with $d$, as the buckling height affects the strength of orbital hybridization, then vanishes at $d > 2.0$ Å where the interaction between the Sn overlayer and the Al substrate becomes negligibly small. On the other hand, because both the Sn adlayer and the Al surface are square lattices, the out-of-plane mirror reflection symmetry ($M_z$) (indicated by the black dashed line in Fig. 1a) is well preserved. Consequently, the gapless feature at the crossing point along $M$ to $X$ is protected. That is to say, a partially broken NL loop is formed in the buckled Lieb lattice of Sn and Al.

Inspired by our theoretical simulation, we carried out eptaxial growth of Sn on an Al(100) substrate. At the early stage of growth, the Sn atoms act as isolated adatoms on Al(100) (Fig. S4). As the coverage of Sn increases (with 1 monolayer (ML) referring to the number of surface Al atoms), Sn adatoms start to coalesce into small islands with height of 1.6 Å, larger than that of the isolated Sn adatoms with 1.3 Å (Fig. 2b and Fig. 2f). This indicates that the Sn−Al interactions within the island are weakened due to the increasing Sn−Sn bond length. When the coverage of Sn reaches 0.5 ML, large-scale flat Sn terraces are formed on the Al(100) surface, with their height of 2.2 Å equal to the step height of the Al(100) (Fig. 2c and 2g), suggesting that the Sn layer has covered the whole surface. The $\sqrt{2} \times \sqrt{2}$ superstructure is verified by both the atomic resolution STM image (Fig. 2d) and the low-energy electron diffraction (LEED) pattern (Fig. 2e). It was also confirmed that the atomic features of the $\sqrt{2} \times \sqrt{2}$ superstructure in STM images are irrelevant to the bias, as shown in Fig. S5. This corroborates the proposal that the atomic resolution STM images indeed display the real-space spatial arrangement of Sn atoms on Al(100). Combining the STM measurements and the coverage, we believe that the adsorbed Sn atoms form a simple $\sqrt{2} \times \sqrt{2}$ adlayer on the Al(100) surface without involving any phase transition.

We performed DFT calculations to determine the exact atomic model of the $\sqrt{2} \times \sqrt{2}$ Sn-Al lattices. After structural optimization, two structural models, the buckled Lieb lattice and a substitutional surface alloy (half of the Al surface replaced by Sn), are found to be much more energetically favorable than any other structure models, as illustrated in Fig. 3a and 3c. Meanwhile, the formation energy of the buckled Lieb lattice mode is about 0.1 eV/atom lower than that of the substitutional surface alloy, suggesting that the former model should be more possible. Other evidence for the buckled Lieb lattice is the trench depth in the $\sqrt{2} \times \sqrt{2}$ structure measured from the STM images shown in Fig. 3e and 3f, which is about 1.8 Å, close to the atom height of Sn. This observation supports the overlayer-structure rather than a surface-alloy, because in a Sn-deficient substitutional surface alloy, a mixed structure of ordered alloy and Al(100) substrate with much smaller height variation than the overlayer (1 Sn atom height) is expected.[28,29] Fig. 3b and 3d show the simulated STM images for both models. It is clear that additional atoms corresponding to Al atoms appear in the center of the square unit cell for the substitutional surface alloy model, which is not observed in our experimental STM images. Therefore, we can conclude that the $\sqrt{2} \times \sqrt{2}$ Sn-Al layer is indeed a 2D buckled Lieb lattice.

Scanning tunneling spectroscopy (STS) and ARPES measurements were then performed to acquire the detailed electronic structure of this buckled Lieb lattice. The $dI/dV$ spectrum shown in Fig. 4a reveals that the $\sqrt{2} \times \sqrt{2}$ surface is metallic, but it shows a largely suppressed local density of states (LDOS) near the Fermi level. In the ARPES measurements, clear surface states (SS) of Al(100)[30-32] were observed but with an apparent up-shift of 0.3 eV towards the Fermi level (Fig. S6). This phenomenon agrees well with the electronegativity difference between Al and Sn: Al atoms with smaller electronegativity donate electrons to top-layer Sn atoms. Accordingly, Mulliken charge analysis shows that each Sn atom receives 0.77 $e$ from the Al substrate and forms strong Sn−Al bonds with bond order of 0.71. The map in the close vicinity of the Fermi surface (Fig. 4b, left panel) shows that the modified surface states intersect with each other and form hole pockets around the $M$ point of the $\sqrt{2} \times \sqrt{2}$ lattice, which results in suppression of the spectral weight of regions near $X$ points. These characteristics of the hole pocket are confirmed by the ARPES map at −700 meV (Fig. 4b, right panel), below which, the intersections of surface states vanished. Since the $M_{Al}$ point of Al(100) and the $\Gamma$ point of the $\sqrt{2} \times \sqrt{2}$ lattice share the same position, the bands near $M_{Al}$ ($\Gamma'$) have been assigned to the bulk band of Al(100) in previous studies.[30] Similarly V bands are likely to be induced by the band-folding effect. In this scenario, V bands near the $\Gamma$ point and the intersected surface states near the $M$ point act as electron and hole pockets, respectively, as they do not overlap with each other at the Fermi surface. These results confirm the metallic character of the buckled Lieb lattice, but with a negative indirect band gap near the Fermi surface.

Furthermore, ARPES intensity plots along the $\Gamma$ to $M$ (along $k_x$) direction are presented in Fig. 4c (Cuts I, II, III and IV) at different $k_x$ positions. Spectral weight suppression of surface states is clearly visible above −0.6 eV near the $X$ point in Cut III. Based on this analysis, a schematic diagram of the Fermi surface in the Brillouin zone (BZ) of the Lieb lattice is given in Fig. 4d, which comprises four hole pockets centered at the $M$ points, an electron pocket centered at the $\Gamma$ point, and suppression of spectral weight near the $X$ point. DFT calculations of the buckled Lieb lattice on bulk Al(100) further demonstrated the orbital hybridization and band folding effect. As shown in Fig. 4e and 4f, the calculated bands well match the ARPES results. The Sn $p_z$ orbitals strongly hybridize with the substrate orbitals and induce up-shifting of the surface states of Al(100) compared with the bare Al substrate (Fig. S7).

From the theoretical point of view, the lack of possible topological insulating states in APRES indicates suppression due to the strong orbital hybridization with the Al(100) substrate. To verify the impact of the substrate on the topological bands, we calculated the band structure of the Sn adlayer with the $\sqrt{2} \times \sqrt{2}$ structure on an Al(100) slab with different numbers of Al layers, as shown in Fig. S8. It is verified that, although the partially broken NL still survives and accompanies stronger orbital hybridization up to 5 layers, it is highly submerged in the valence states of the substrate with increasing thickness of Al(100) slab, and would be hard to detect by either STS or ARPES. Thus, in order to retain such topological fermions, it is necessary to fabricate this buckled Lieb lattice on a substrate with only a few layers of Al atoms or an insulating substrate.

In summary, we have demonstrated a feasible way to design and fabricate a realistic 2D Lieb lattice composed of an Sn overlayer on Al(100) substrate by the molecular beam epitaxy (MBE) method. Suitable hybridization interaction and charge transfer between the Sn sublattice and the substrate are vital for stabilizing this unique square lattice structure. Suppression of spectral weight of surface states in STS and ARPES have been measured due to the existence of this buckled lieb lattice on Al(100). The Topological insulating states protected by its mirror reflection symmetry have been predicted in this superlattice. The experimental confirmation of 2D Lieb lattices would promote our understanding of their symmetry-related electronic and topological properties for the exploration of novel physical properties and applications.

**Experimental sections**

The Al(100) substrate (Mateck) was purchased from Mateck, GmbH. The surface was prepared by cycles (more than 10) of Ar$^+$ sputtering (1 kV, 30 min) and annealing (450 °C, 10 min). Sn was deposited on the cleaned Al(100) surface by evaporation of Sn from a home-built Ta crucible at room temperature. The growth rate of Sn was calibrated by the growth of Sn on Ag(111) substrate, which forms a $\sqrt{3} \times \sqrt{3}$ surface reconstruction at the coverage of 1/3 ML. STM was performed by using a low-temperature (LT) ultrahigh-vacuum (UHV) scanning tunnelling microscope from Unisoku Co. (Unisoku 1400) at 80 K. All the STM images were obtained in constant current mode. The STS differential conductance ($dI/dV$) was obtained with lock-in detection by applying a small modulation to the tunnel voltage at 973 Hz. The STM images were analysed using WSxM software.[33] In-situ LEED and ARPES characterizations (SCIENTA R4000 analyzer) were carried out at Beamline 4B9B in the Beijing Synchrotron Radiation Facility (BSRF). The sample was prepared by the same method as in the STM study. All the data were recorded in UHV at room temperature.

DFT calculations were performed with the Vienna ab initio simulation package, using the plane wave basis set with an energy cut-off of 500 eV,[34] projector augmented wave potentials,[35] and the generalized gradient approximation parameterized by Perdew, Burke and Ernzerhof (PBE) for the exchange-correlation functional.[36] A 15-atomic-layer slab model was adopted for the Al(100) substrate, which, according to our test calculations, is necessary for obtaining electronic band structures consistent with the ARPES results for Sn overlayers on Al(100). The supercell consists of a Sn atom on $\sqrt{2} \times \sqrt{2}$ unit cells of Al(100), with a vacuum region of 15 Å in the vertical direction. The Brillouin zone was sampled by a $12 \times 12 \times 1$ uniform k-point mesh. The model structures were fully optimized for the electronic and cell degrees of freedom with thresholds of $10^{-6}$ eV for the total energy and $10^{-3}$ eV/Å for the force on each atom. The binding energy of Sn atoms on Al(100) is defined as $\Delta E_{\mathrm{b}} = E_{\mathrm{Sn}} + E_{\mathrm{sub}} - E_{\mathrm{Sn+sub}}$, where $E_{\mathrm{Sn+sub}}$ and $E_{\mathrm{sub}}$ are the energies of Al(100) with and without adsorption of an Sn atom, respectively, and $E_{\mathrm{Sn}}$ is the energy of a single Sn atom. The STM images were simulated by the Tersoff-Hamann approximation.[37] The Z2 topological invariant was calculated by constructing the Wannier functions and establishing the evolution of hybrid Wannier charge centers and their largest gap function, with SOC considered.

**Figures and figure captions**

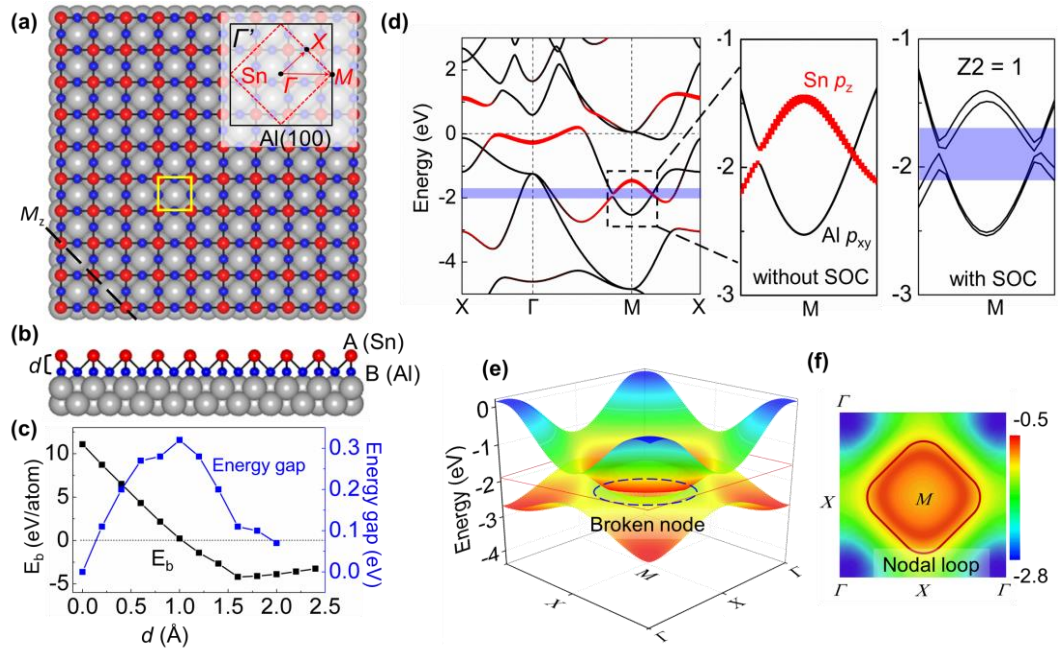

**Figure 1.** Top view (a) and side view (b) of the buckled Lieb lattice comprising Sn in a $\sqrt{2} \times \sqrt{2}$ superstructure and the topmost Al atoms of the Al(100) substrate. The red balls and blue (gray) balls represent Sn and the surface (underlying) Al atoms, respectively. The unit cell is indicated by the yellow square. The Brillouin zone (BZ) is shown as the inset in (a). (c) Calculated binding energy and energy gap of the broken NL of the buckled Lieb lattice as a function of the interlayer distance ($d$). (d) Calculated band structure of the buckled Lieb lattice with an optimal interlayer distance of 1.6 Å, with zoom-in plots of the NL with/without SOC shown in the right panels. The Fermi energy is shifted to zero. The partial band projection from the $p_z$ orbitals of Sn atoms is given by the red symbols, with the symbol size proportional to the band weight. (e) 3D band structure, with the black (red) dashed line indicating (the plane of) the broken node. (f) Cross-section of 3D band structure at $E = -1.9$ eV, showing a square NL loop.

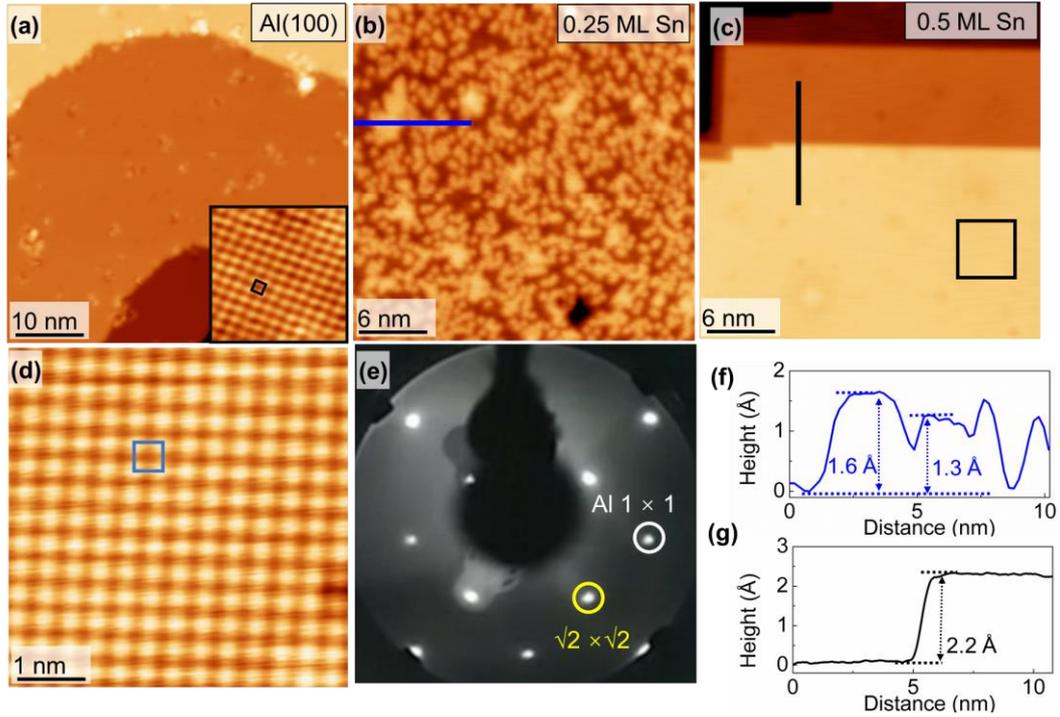

**Figure 2.** (a) STM image of the Al(100) substrate (sample bias $V_S = 2$ V, tunneling current $I = 50$ pA), where the inset is the atomic resolution image of Al(100), with the unit cell marked by the black square. (b) STM image of 0.25 ML Sn on the Al(100) surface ($V_S = 1.5$ V, $I = 50$ pA). (c) STM image of 0.5 ML Sn on Al(100) ($V_S = 1.5$ V, $I = 50$ pA). (d) Atomic resolution STM image of 0.5 ML Sn on Al(100) ($V_S = 70$ mV, $I = 50$ pA), showing a square lattice with lattice constant of 0.4 nm, as indicated by the blue square. (e) LEED pattern of Sn overlayer on Al(100), with the $\sqrt{2} \times \sqrt{2}$ structure and the $1 \times 1$ diffraction pattern indicated by the white and yellow circles, respectively. (f) Height profile along the blue line in (b). (g) Height profile along the black line in (c).

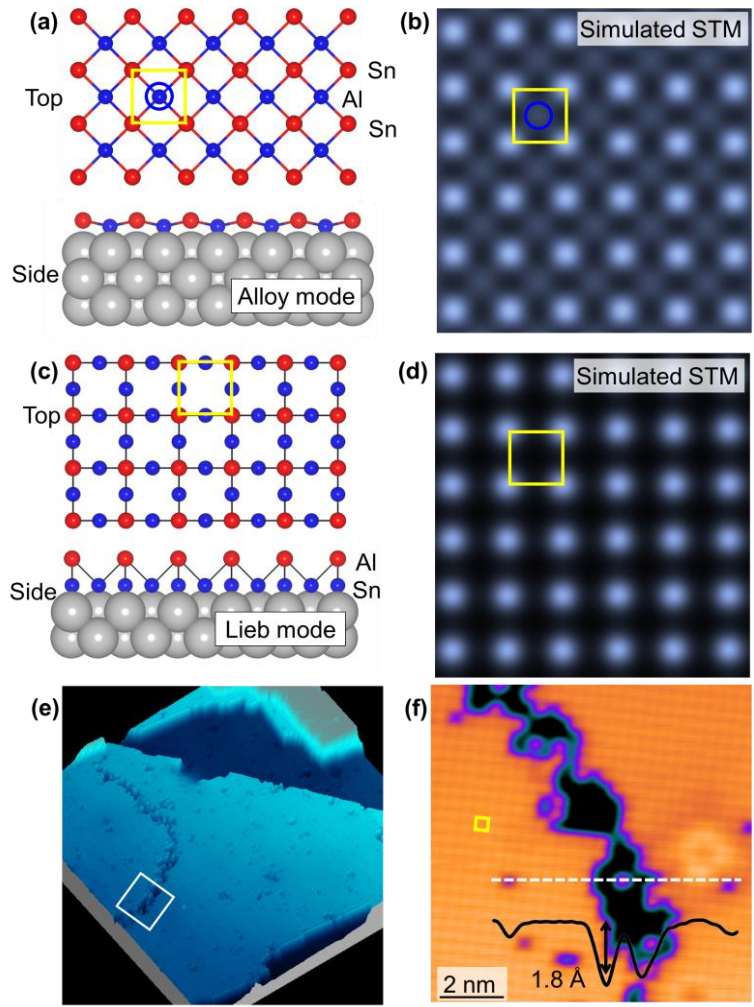

**Figure 3.** (a) Structure of a Sn-deficient Sn $\sqrt{2} \times \sqrt{2}$ superstructure on Al(100), with abundant defects on the terrace. (b) Enlarged STM image of an area with the bare Al(100) surface exposed. (c) Structure of SnAl alloy on Al(100). (d) Simulated STM image based on the structural model in (c). (e) Structure of the buckled Lieb lattice on Al(100). (f) Simulated STM image based on the structural model in (e).

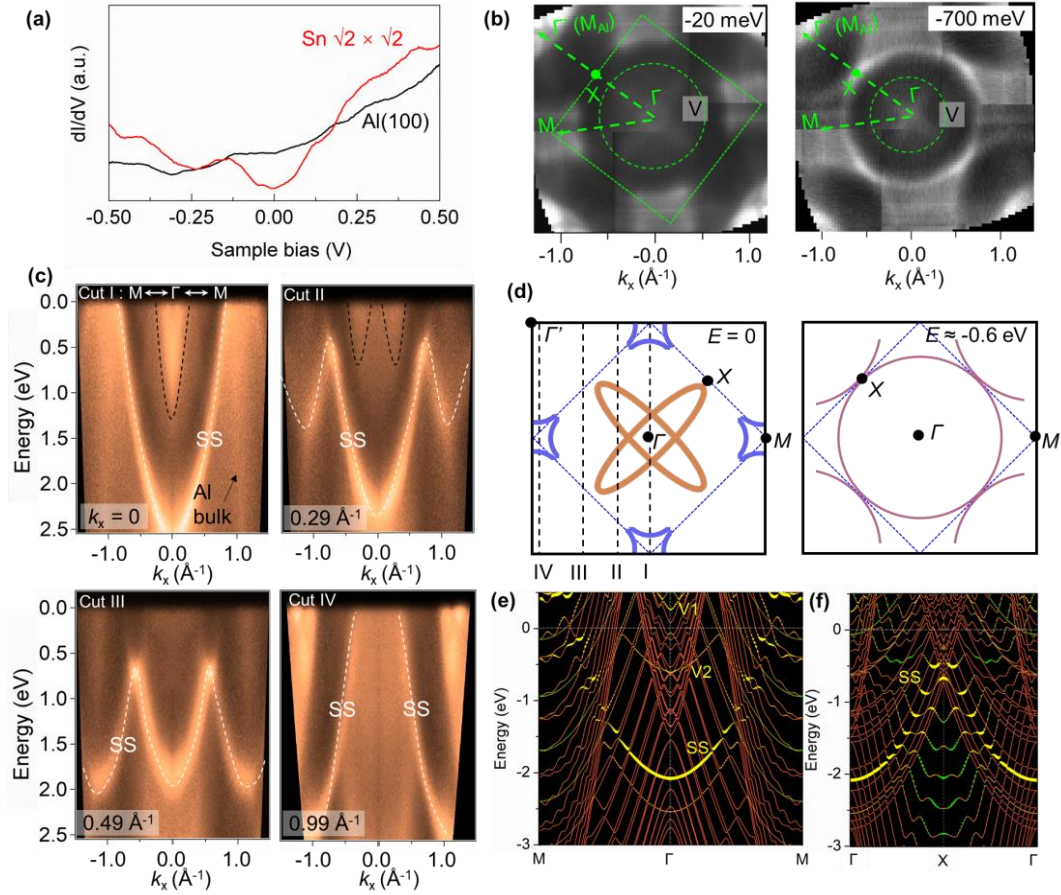

**Figure 4.** (a) *dI/dV* spectrum of Sn √2 × √2 superstructure on Al(100), along with the *dI/dV* spectrum of bare Al(100). (b) Fermi surface mapping of Sn √2 × √2 on Al(100) at *E* = −20 meV and *E* = −700 meV, with the 2D BZ indicated by the green dotted square. (c) ARPES intensity plot along the *Γ* to *M* direction at $k_x = 0$, 0.29 Å⁻¹, 0.49 Å⁻¹, and 0.99 Å⁻¹. (d) Schematic diagram of surface state (SS) and states of bands A at the Fermi level and at 0.6 eV below the Fermi energy, with the blue and brown colors representing the hole and electron pockets, respectively (2D BZ represented by blue dotted square). (e), (f) Calculated band structures of Sn √2 × √2 on Al(100) (modelled by a slab model with 15 atomic layers) along different directions of the BZ, with the partial band projected on the $p_{xy}$ and $p_z$ orbitals of Sn atoms, as indicated by green and yellow colors, respectively.